\documentclass{elsart}

\usepackage{amsbsy,amssymb}

\usepackage{subfigure}
\usepackage{epsfig}

\def\vec#1{{\boldsymbol #1}}

\providecommand\boldsymbol[1]{\mbox{\boldmath $##1$}}

\begin{document}

\begin{frontmatter}



\title{Breakdown of Burton-Prim-Slichter approach and lateral solute
segregation in radially converging flows}

\author[IPUL,FZR]{J\={a}nis Priede},
\ead{priede@sal.lv}
\author[FZR]{Gunter Gerbeth\corauthref{cor}}
\ead{g.gerbeth@fz-rossendorf.de}
\address[IPUL]{Institute of Physics, University of Latvia, LV-2169 Salaspils, Latvia}
\address[FZR]{Forschungszentrum Rossendorf, MHD Department, PO Box 510119, 01314 Dresden, Germany}

\corauth[cor]{Corresponding author. Tel. +49-351-2603484; fax:
+49-351-2602007.}

\date{18 August 2005}

\begin{abstract}
A theoretical study is presented of the effect of a radially
converging melt flow, which is directed away from the
solidification front, on the radial solute segregation in simple
solidification models. We show that the classical
Burton-Prim-Slichter (BPS) solution describing the effect of a
diverging flow on the solute incorporation into the solidifying
material breaks down for the flows converging along the
solidification front. The breakdown is caused by a divergence of
the integral defining the effective boundary layer thickness which
is the basic concept of the BPS theory. Although such a divergence
can formally be avoided by restricting the axial extension of the
melt to a layer of finite height, radially uniform solute
distributions are possible only for weak melt flows with an axial
velocity away from the solidification front comparable to the
growth rate. There is a critical melt velocity for each growth
rate at which the solution passes through a singularity and
becomes physically inconsistent for stronger melt flows. To
resolve these inconsistencies we consider a solidification front
presented by a disk of finite radius $R_{0}$ subject to a strong
converging melt flow and obtain an analytic solution showing that
the radial solute concentration depends on the radius $r$ as
$\sim\ln^{1/3}(R_{0}/r)$ and $\sim\ln(R_{0}/r)$ close to the rim
and at large distances from it. The logarithmic increase of
concentration is limited in the vicinity of the symmetry axis
by the diffusion becoming effective at a distance comparable to
the characteristic thickness of the solute boundary layer.
The converging flow causes a solute pile-up forming a
logarithmic concentration peak at the symmetry axis which might be
an undesirable feature for crystal growth processes.
\end{abstract}

\end{frontmatter}


\section{Introduction}

Solidification and crystallisation processes are present in
various natural phenomena as well as in a large number of material
production technologies such as, for example, semiconductor
crystal growth from the melt, alloy metallurgy, etc. Usually the
melt used for the production of solid material is not a pure
substance but rather a solution containing some dissolved dopants
or impurities. Often the solid material grown from the solution
has a non-uniform distribution of the dissolved substance although
the original solution was uniform. This non-uniformity is caused
by the difference of equilibrium concentrations of solute in the
liquid and solid phases. Thus, if the equilibrium concentration of
solute in a crystal is lower than in the melt, only a fraction of
solute is incorporated from the melt into the growing crystal
while the remaining part is repelled by the solidification front
as it advances into the liquid phase \cite{Hurle}. This effect
causes axial segregation of the solute, usually concentrated in a
thin, diffusion-controlled boundary layer adjacent to the
solidification front. Axial segregation can strongly be influenced
by the melt convection. According to the original work by Burton,
Prim and Slichter (BPS) \cite{BPS}, a sufficiently strong
convection towards the crystallisation front reduces the thickness
of the segregation boundary layer and so the solute concentration
getting into the crystal. Such a concept of solute boundary layer
has been widely accepted to interpret the effect of melt flow on
the solute distribution in various crystal growth configurations
\cite{CamelFavier83,CamelFavier86,Garandetetal}.
The BPS approach, originally devised for a rotating-disk flow
modelling an idealised Czochralski growth configuration, supposes
the melt to be driven towards the solidification front by a
radially diverging flow. However, in many cases, as for instance
in a flow rotating over a disk at rest \cite{Schlichting}, like in
a flow driven by a rotating \cite{Davidson} or a travelling
\cite{Yesil04} magnetic field, as well as in the natural
convection above a concave solidification front in the vertical
Bridgman growth process \cite{ChangBrown83}, the melt is driven
away from the solidification front in its central part by a
radially converging flow. Though several extensions of the BPS
solution exist (e.g. \cite{Wilson78,Wheeler80,HurleSeries,Cartwright}),
the possibility of a reversed flow direction away from the
crystallisation front has not yet been considered in that context.

In this work, we show that the BPS approach becomes invalid for
converging flows because the effective boundary layer thickness,
which is the basic concept of the BPS theory, is defined by an
integral diverging for a flow away from the solidification front.
The divergence can formally be avoided by restricting the space
occupied by the melt above the solidification front to a layer of
finite depth, but for higher melt velocities this solution becomes
physically inconsistent, too. Next we consider a solidification
front as a disk of finite radius immersed in the melt with a
strong converging flow and show that a converging flow results in
a logarithmic solute segregation along the solidification front
with a peak at the symmetry axis. An analytical solution is
obtained by an original technique using a Laplace transform. The
advantage of this solution is its simple analytical form as well
as the high accuracy which has been verified by comparing with
numerical solutions.

The simulation of dopant transport is an important aspect of
crystal growth modelling \cite{Hirtz,Lan}, and various
numerical approaches are used for it. However, a numerical
approach is always limited in the sense that it provides only
particular solutions while the basic relations may remain hidden.
Besides, the numerical solution often requires considerable
computer resources when a high spatial resolution is necessary
which is particularly the case for thin solute boundary layers. It
has been shown, \textit{e.g.}, by Vartak and Derby \cite{Vartak}
that an insufficient resolution of the solute boundary layer may
lead to numerically converged but nevertheless inaccurate results.

The paper is organised as follows. In Section 2 we discuss the
BPS-type approach and show its inapplicability to converging
flows. The simple model problem of radial segregation along a disk
of finite radius in a strong converging flow is described in
Section 3, and an analytical solution for the concentration
distribution on the disk surface is obtained in Section 4. Summary
and conclusions are presented in Section 5.

\section{Breakdown of BPS-type solutions}

Consider a simple solidification model consisting of a flat
radially-unbounded solidification front advancing at velocity
$v_{0}$ into a half-space occupied by the melt which is a dilute
solution characterised by the solute concentration $C.$ The latter
is assumed to be uniform and equal to $C_{\infty}$ sufficiently
far away from the solidification front. Solute is transported in
the melt by both diffusion with a coefficient $D$ and the melt
convection with a velocity field $\vec{v}$. At the solidification
front, supposed to be at the thermodynamic equilibrium, the ratio
of solute concentrations in the solid and liquid phases is given
by the equilibrium partition coefficient $k$. In the absence of
convection, the repelled solute concentrates in a boundary layer
with the characteristic thickness $\delta_{0}=D/v_{0}$. We
consider in the following the usual case of a much larger momentum
boundary layer compared to the solute boundary layer, \textit{i.e.} a high
Schmidt number $\mathit{Sc}=\nu/D\gg1$ where $\nu$ is the kinematic
viscosity of the melt. The basic assumption of the BPS approach is
that the lateral segregation is negligible and thus the solute
transport is affected only by the normal velocity component. The
latter is approximated in the solute boundary layer by a power
series expansion in the distance $z$ from the solidification front
as $v(z)\approx\frac{1}{2}v''(0)z^{2}.$ Then the equation
governing the concentration distribution in the solute boundary
layer may be written in dimensionless form as

\begin{equation}
-(1+\mathit{Pe}z^{2})\frac{dC}{dz}=\frac{d^{2}C}{dz^{2}},
\label{eq:BPS}
\end{equation}
where $\mathit{Pe}=\frac{v''(0)\delta_{0}^{3}}{2D}$ is the local
P\'{e}clet number based on the characteristic boundary layer
thickness $\delta_{0}$ which is used as length scale here while
the concentration is scaled by $C_{\infty}.$ The boundary
conditions for the uniformly mixed melt core and the solid-liquid
interface take the form $\left.C\right|_{z\rightarrow\infty}\rightarrow1$
and
\begin{equation}
\left[(1-k)C+\frac{dC}{dz}\right]_{z=0}=0.
\label{BPS.bnd}
\end{equation}
The solution of this problem is
\begin{equation}
C(z)=1+A\int_{z}^{\infty}\exp\left(-t-\frac{\mathit{Pe}}{3}t^{3}\right)\,dt,
\label{BPS.sol}
\end{equation}
where the constant $A=\frac{1-k}{1-(1-k)\Delta(\mathit{Pe})}$ is
obtained from (\ref{BPS.bnd}) in terms of $\Delta(\mathit{Pe})=
\int_{0}^{\infty}\exp\left(-t-\frac{\mathit{Pe}}{3}t^{3}\right)\,dt$ which
according to the relation
$C'(0)=\frac{C(\infty)-C(0)}{\Delta(\mathit{Pe})}$ represents an
effective dimensionless thickness of the solute boundary layer.
Eventually, the concentration at the solidification front is
obtained as $C(0)=\left[1-(1-k)\Delta(\mathit{Pe})\right]^{-1}$.
This is the central result of the BPS approach stating that only
the effective thickness of the solute boundary layer defined by
the local velocity profile is necessary to find the solute
concentration at the solidification front for a given uniform
concentration in the bulk of the melt. However, it is important to
note that this solution is limited to $\mathit{Pe}\geq0$ and it
becomes invalid for $\mathit{Pe}<0$ when the flow is directed away
from the solidification front because both integrals in Eq.
(\ref{BPS.sol}) and $\Delta(\mathit{Pe})$ diverge in this case.
The goal of this study is to find out what happens to the solute
distribution when the flow is directed away from the
solidification front and the BPS solution breaks down.

\begin{figure}
\centering
\includegraphics[width=0.75\columnwidth]{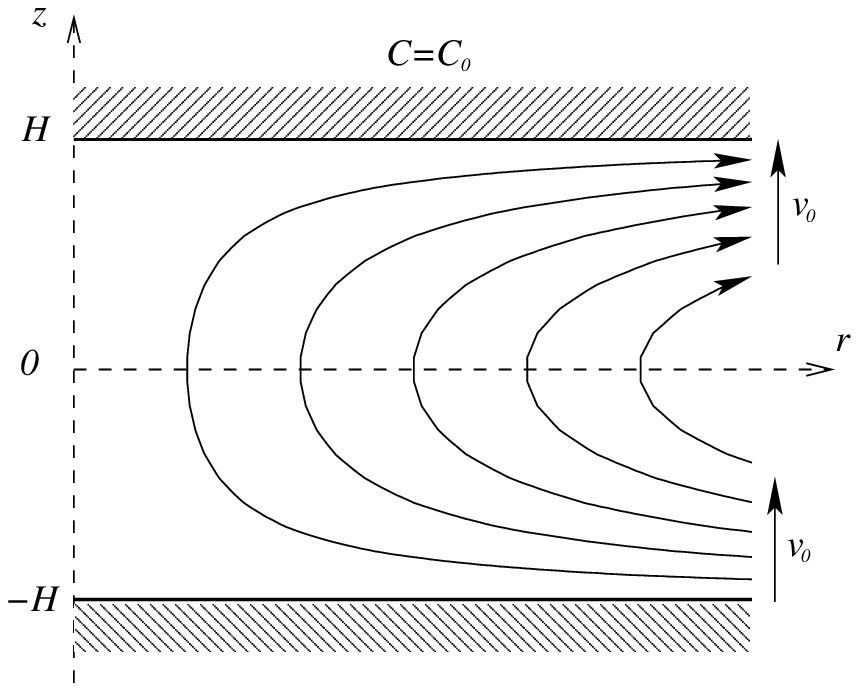}

\caption{
\label{cap:sketch1}
Sketch of a radially unbounded flat layer with solidification and
melting fronts at bottom and top, respectively. }
\end{figure}

The divergence in the BPS model for $\mathit{Pe}<0$ is obviously
related to the unbounded interval of integration which can be
avoided by taking into account the finite axial size of the
system. The simplest such model, shown in Fig. \ref{cap:sketch1},
is provided by a flat, radially-unbounded layer between two disks
separated by a distance $2H$. The upper and lower disks represent
the melting and solidification fronts, respectively, and the
molten zone proceeds upwards with velocity $v_{0}.$ There is a
forced convection in the melt with the axial velocity $v(z)$ which
is assumed to satisfy impermeability and no-slip boundary
conditions. There is also a radial velocity component following
from the incompressibility constraint which, however, is not
relevant as long as a radially uniform concentration distribution
is considered. Here we choose $H$ as a length scale so that the
boundaries are at $z=\pm1$. At the upper boundary, there is a
constant solute flux due to the melting of the feed rod with the
given uniform concentration $C_{0}$ with velocity $v_{0}$
\[
\left.\mathit{Pe}_{0}(C-C_{0})+\frac{dC}{dz}\right|_{z=1}=0.
\]
Note that this boundary condition following from the mass conservation
does not formally satisfy the local thermodynamic equilibrium
relating the solute concentrations in the solid and liquid phases.
In order to ensure equilibrium concentrations at the melting front
it would be necessary to take into account also the diffusion in
the solid phase which, however, is neglected here. Such an
approximation is justified by the smallness of the corresponding
diffusion coefficient.

\begin{figure}
\centering
\includegraphics[width=0.75\columnwidth]{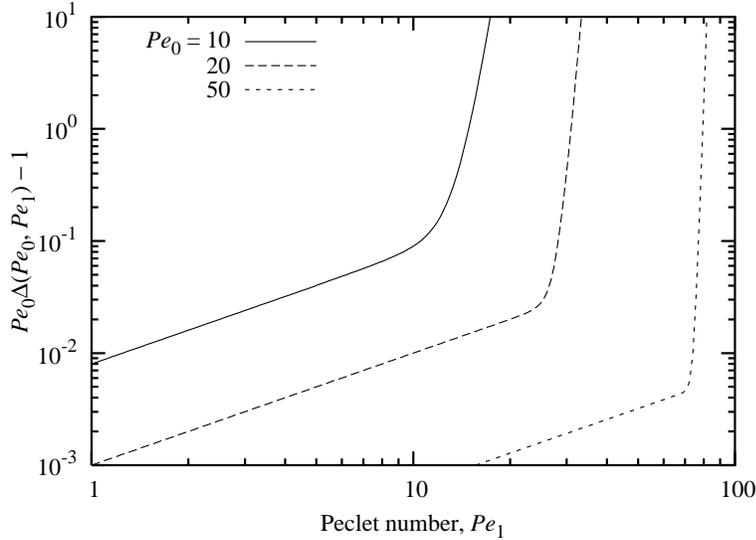}

\caption{ \label{cap:BPSm} Modified effective boundary layer
thickness $\mathit{Pe}_{0}\Delta(\mathit{Pe}_{0},\mathit{Pe}_{1})-1$
at the solidification front for a horizontal liquid layer of finite
height with the flow away from the solidification front versus the
P\'{e}clet number $\mathit{Pe}_{1}$ of melt stirring at various
P\'{e}clet numbers $\mathit{Pe}_{0}$ based on the solidification
rate.}
\end{figure}

At the lower boundary, coinciding with the moving solidification
front, the boundary condition is
\begin{equation}
\left.(1-k)\mathit{Pe}_{0}C+\frac{dC}{dz}\right|_{z=-1}=0,
\label{bnd:BPS-m}
\end{equation}
where $\mathit{Pe}_{0}=v_{0}H/D$ is the P\'{e}clet number based
on the solidification velocity. The radially uniform concentration
distribution depending only on the axial coordinate $z$ is
governed by
\begin{equation}
\left(-\mathit{Pe}_{0}+\mathit{Pe}_{1}v(z)\right)
\frac{dC}{dz}=\frac{d^{2}C}{dz^{2}},
\label{eq:BPS-m}
\end{equation}
where $\mathit{Pe}_{1}$ is the P\'{e}clet number of convection.
The solution of the above equation is
\begin{equation}
C(z)=A+B\int_{-1}^{z}\exp\left[-\mathit{Pe}_{0}(t+1)
+\mathit{Pe}_{1}\int_{-1}^{t}v(\tau)\,d\tau\right]\, dt.
\label{sol:BPS-m}
\end{equation}
The boundary condition (\ref{bnd:BPS-m}) yields $B=-A(1-k)\mathit{Pe}_{0}$
while the remaining unknown constant $A$ is determined from the
condition at the upper boundary. However, for our purposes it is
sufficient to express $A$ in terms of the concentration at the
solidification front: $A=C(-1).$ Then Eq. (\ref{sol:BPS-m}) allows
us to relate the concentrations at the melting and solidification
fronts
\begin{equation}
C(-1)=C(1)\left[1-(1-k)\mathit{Pe}_{0}
\Delta(\mathit{Pe}_{0},\mathit{Pe}_{1})\right]^{-1},
\label{c0-m}
\end{equation}
where
\begin{equation}
\Delta(\mathit{Pe}_{0},\mathit{Pe}_{1})=\int_{-1}^{1}\exp
\left[-\mathit{Pe}_{0}(t+1)+\mathit{Pe}_{1}\int_{-1}^{t}v(\tau)\,d\tau\right]\,dt
\label{delta-gen.mod}
\end{equation}
is the effective solute boundary layer thickness defined by the
relation $\left.\frac{dC}{dz}\right|_{z=-1}=\frac{C(1)-C(-1)}
{\Delta(\mathit{Pe}_{0},\mathit{Pe}_{1})}$ following from Eqs.
(\ref{bnd:BPS-m}) and (\ref{c0-m}). This effective boundary layer
thickness at the solidification front is plotted in Fig.
\ref{cap:BPSm} for a model velocity distribution
$v(z)=\left(1-z^{2}\right)^{2}$. The effective boundary layer
thickness increases with the convection but the increase is
relatively weak until $\mathit{Pe}_{1}$ becomes comparable to
$\mathit{Pe}_{0}$. At this point, the effective thickness starts
to grow nearly exponentially.

Although the effective boundary layer thickness is now bounded for
any finite value of $\mathit{Pe}_{1}$ regardless of its sign
defining the flow direction, the obtained solution is not really
free from singularities. At first, note that the concentration at
the solid-liquid interface becomes singular when the solute
boundary layer becomes as thick as
$\mathit{Pe}_{0}\Delta(\mathit{Pe}_{0},\mathit{Pe}_{1})=(1-k)^{-1}$
resulting in a zero denominator in Eq. (\ref{c0-m}). Second, for
larger $\mathit{Pe}_{1}$ the denominator in Eq. (\ref{c0-m})
becomes negative implying a negative concentration at the
solidification front that presents an obvious physical
inconsistency. Thus, the obtained solution is applicable only for
sufficiently weak converging flows and breaks down as the velocity
of the melt flow away from the solidification front becomes
comparable to the growth rate at $\mathit{Pe}_{1}\sim\mathit{Pe}_{0}.$

\section{A disk of finite radius with a strong converging flow}

\begin{figure}
\centering
\includegraphics[width=0.75\columnwidth]{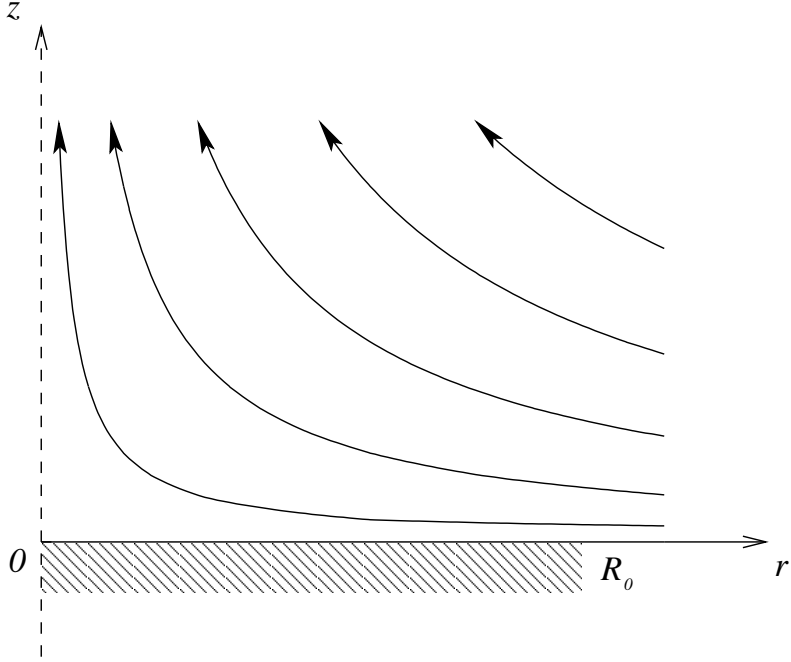}

\caption{
\label{cap:sketch2}
Sketch of the solidification front presented by a disk of radius
$R_{0}$ in a converging flow.}
\end{figure}

The assumption underlying both the classical BPS approach and that
of the previous section is the neglected radial segregation. The
simplest physical model which could account for radial segregation
is presented by a solidification front in the form of a disk of
finite radius $R_{0}$ with the melt occupying the half-space above
it, as shown in Fig. \ref{cap:sketch2}. For simplicity, the
velocity distribution in the melt is assumed to be that of a
liquid rotating above a disk at rest. In this case, contrary to
the classical BPS problem of a rotating disk, the flow is radially
converging rather than diverging. Thus, within the solute boundary
layer, assumed as usual to be thin relative to the momentum
boundary layer, the radial and axial velocity components can be
approximated as
\[
v_{r}\approx-\frac{1}{2}v_{z}''(0)rz, \qquad
v_{z}\approx\frac{1}{2}v_{z}''(0)z^{2}.
\]
Here we choose the thickness of the solute boundary layer based on
the axial melt velocity as length scale
\begin{equation}
d_{0}=(2D/v_{z}''(0))^{1/3},
\label{eq:d0}
\end{equation}
and assume the stirring of the melt to be so strong that the advancement
of the solidification front with the growth velocity $v_{0}$ is
small compared the characteristic melt flow in the solute boundary
layer. The last assumption implies that the local P\'{e}clet
number based on the growth rate is small:
$\tilde{\mathit{Pe}}_{0}=v_{0}d_{0}/D\ll1.$ Then the problem is defined by
a single dimensionless parameter, the dimensionless radius
$R=R_{0}/d_{0}=R_{0}(2D/v_{z}''(0))^{-1/3},$ which may be regarded
as P\'{e}clet number based on the external length scale $R_{0}$
and the internal velocity scale $v_{0}=v_{z}''(0)d_{0}^{2}/2.$ The
governing dimensionless equation is
\begin{equation}
z\left(z\frac{\partial C}{\partial z}-r\frac{\partial C}{\partial r}\right)
=\frac{1}{r}\frac{\partial}{\partial r}
\left(r\frac{\partial C}{\partial r}\right)
+\frac{\partial^{2}C}{\partial z^{2}},
\label{eq:C-rbnd}
\end{equation}
where the radial diffusion term will be neglected as usual for the
boundary layer solution to be obtained in the following.
Sufficiently far away from the solidification front a well-mixed
melt is assumed with a uniform dimensionless concentration
$C_{0}=1.$ The boundary condition at the solidification front
\[
\left.\tilde{\mathit{Pe}}_{0}(1-k)C+\frac{\partial C}
{\partial z}\right|_{z=0}=0,
\]
for $\tilde{\mathit{Pe}}_{0}\ll1$ suggests to search for the concentration as
\begin{equation} C\approx C_{0}+\tilde{\mathit{Pe}}_{0}(1-k)C_{1},
\label{eq:C1def}
\end{equation}
where $C_{1}$ is the deviation of the concentration with a
characteristic magnitude $\tilde{\mathit{Pe}}_{0}(1-k)\ll1$ from its
uniform core value $C_{0}=1.$ Then the boundary condition for
$C_{1}$ takes the form $\left.\frac{\partial C_{1}}{\partial
z}\right|_{z=0}=-1,$ while $C$ is substituted by $C_{1}$ in Eq.
(\ref{eq:C-rbnd}) which, compared to the original BPS Eq.
(\ref{eq:BPS}), has an extra term related to the radial advection
whereas both the term of axial advection due to the solidification speed
and the radial diffusion term have been neglected.
Note that on one hand the radial advection
term is indeed important because without it we recover the BPS
case which was shown above to have no bounded solution. On the
other hand, for the radial advection term to be significant the
solute distribution has to be radially nonuniform. However,
searching for a self-similar solution in the form
$C_{1}(r,z)=r^{\alpha}F(zr^{\beta})$ leads only to the radially
uniform solution with $\alpha=\beta=0$. This implies that a
possible solution has to incorporate the radial length scale $R$.
Additional difficulties with finding similarity solutions are
caused by the explicit appearance of $r$ in Eq. (\ref{eq:C-rbnd}).
Both these facts suggest the substitution $\tau=-\ln(r)$ that
transforms Eq. (\ref{eq:C-rbnd}) into
\begin{equation}
z\left(z\frac{\partial C}{\partial z} +\frac{\partial
C}{\partial\tau}\right) =\frac{\partial^{2}C}{\partial z^{2}}
\label{eq:C-tau}
\end{equation}
with the radial diffusion term neglected as mentioned above. Since
the transformed equation does not explicitly contain $\tau,$
$C(\tau,z)$ being a solution implies that $C(\tau-\tau_{0},z)$ is
also a solution. Consequently, $\tau$ can be replaced by
$\tau-\tau_{0},$ where $\tau_{0}=-\ln(R)$ and thus
$\tau=\ln(R/r)$. Note that $\tau=0$ corresponds to the rim of the
disk while $\tau\rightarrow\infty$ to the symmetry axis.

\section{Solution by Laplace transform}

Equation (\ref{eq:C-tau}) can efficiently be solved by a Laplace
transform providing asymptotic solutions of the solute
distribution along the solidification front for both small and
large $\tau$. The Laplace transform defined as
$\bar{C}(s,z)=\int_{0}^{\infty}C_{1}(\tau,z)e^{-s\tau}d\tau$
transforms Eq. (\ref{eq:C-tau}) into
\[
z\left(z\frac{d\bar{C}}{dz}+s\bar{C}\right)=\frac{d^{2}\bar{C}}{dz^{2}},
\]
where $s$ is a complex transformation parameter while the boundary
condition at the solidification front takes the form:
$\left.\frac{\partial\bar{C}}{\partial
z}\right|_{z=0}=-\frac{1}{s}.$ A bounded solution of this problem
is
$\bar{C}(s,z)=cU\left(\frac{s}{3},\frac{2}{3},\frac{z^{3}}{3}\right),$
where $U(a,b,x)$ is the confluent hypergeometric function
\cite{Abramowitz}. The constant $c$ is determined from the
boundary condition at the solidification front as
$c=\frac{3^{-2/3}}{s}\frac{\Gamma(s/3)}{\Gamma(2/3)}$. At the
solidification front we obtain
\[
\bar{C}(s,0)=\frac{3^{-2/3}}{s}\frac{\Gamma(1/3)}{\Gamma(2/3)}
F\left(\frac{s}{3};\frac{1}{3}\right),
\]
where
\begin{equation}
F(p;a)=\frac{\Gamma(p)}{\Gamma(p+a)}.
\label{eq:ffun}
\end{equation}
The concentration distribution along the solidification front is then given
by the inverse Laplace transform
\[
C_{1}(\tau,0)=\frac{1}{2\pi
i}\int_{b-i\infty}^{b+i\infty}e^{st}\bar{C}(s,0)ds.
\]
The solution for small $\tau$ follows from the asymptotic
expansion of $F(p;a)$ at $\left|p\right|\gg1$ that can be
presented as
\[
F\left(\frac{s}{3};\frac{1}{3}\right)=
\sum_{j=0}f_{j}\left(\frac{1}{3}\right)\left(\frac{s}{3}\right)^{-j-1/3},
\]
where $f_{j}(a)$ are the asymptotic expansion coefficients of
$F(p;a)=p^{-a}\sum_{j=0}\frac{f_{j}(a)}{p^{j}}$
which can be found efficiently by the following approach. We start
with the basic relation $F(p;a)=(1+a/p)F(p+1;a)$ resulting from
(\ref{eq:ffun}). The asymptotic expansion of both sides of this relation
can be presented as
\begin{equation}
\sum_{j=0}\frac{f_{j}(a)}{p^{j}}=\sum_{j=0}
\frac{f_{j}(a)}{p^{j}}g_{j}(p;a),
\label{eq:renorm}
\end{equation}
where $g_{j}(p;a)=\left(1+ap^{-1}\right)\left(1+p^{-1}\right)^{a-j}
=\sum_{l=0}\frac{g_{j,l}(a)}{p^{l}}$
with the expansion coefficients
\[
g_{j,l}(a)=
\left\{
\begin{array}{cc}
 1, & l=0\\
\frac{(-1)^{l}}{l!}(a+j)_{l-1}(j+(l-1)(1-a)), & l>0
\end{array}
\right.,
\]
defined by use of Pochhammer's symbol $(p)_{n}=\frac{\Gamma(p+n)}{\Gamma(p)}$.
Substituting the above expansion back into Eq. (\ref{eq:renorm}) and comparing
the terms with equal powers of $p$ we obtain
$f_{j}(a)=\sum_{l=0}^{j}f_{l}(a)g_{l,j-l}(a),$ that due to
$g_{l,0}=1$ simplifies to $\sum_{l=0}^{j-1}f_{l}(a)g_{l,j-l}(a)=0.$
Upon replacing $j$ by $j+1$ and taking into account $g_{j,1}(a)=-j$, the latter
relation results in
\begin{equation}
f_{j}(a)=\frac{1}{j}\sum_{l=0}^{j-1}f_{l}(a)g_{l,j+1-l}(a),
\label{eq:recurs}
\end{equation}
defining $f_{j}(a)$ recursively for $j>0.$ In order to apply this
recursion we need $f_{0}(a)$ which can be shown to be constant and
therefore $f_{0}(a)=1$ because $f_{0}(0)=1.$ Eventually, we obtain
\begin{equation}
C_{1}(\tau,0)=\frac{3^{2/3}}{\Gamma(2/3)}\sum_{j=0}d_{j}\tau^{j+1/3},
\label{sol:powser}
\end{equation}
where $d_{j}=\frac{3^{j-1}f_{j}(1/3)}{(1/3)_{j+1}}$. That means,
the radial solute segregation along the solidification front at
the rim is characterised by the leading term
$C_{1}(r,0)\approx\frac{3^{2/3}}{\Gamma(2/3)}\ln^{1/3}(R/r)$. The
first 9 coefficients of the series expansion (\ref{sol:powser})
calculated analytically by \textit{Mathematica} \cite{Mathematica}
are shown in Table \ref{cap:table}. The convergence of the
obtained power-series solution is limited to
$\tau\leq\lim_{j\rightarrow\infty}\sqrt{-\frac{d_{j}}{d_{j+2}}}\approx2.09$
\cite{Hinch}.

\begin{table}
\begin{center}\begin{tabular}{|c|c|c|c|c|c|c|c|c|c|}
\hline $j$&
 $0$&
 $1$&
 $2$&
 $3$&
 $4$&
 $5$&
 $6$&
 $7$&
 $8$\tabularnewline
\hline $d_{j}$&
 $1$&
 $\frac{1}{4}$&
 $\frac{1}{28}$&
 $-\frac{1}{120}$&
 $-\frac{1}{390}$&
 $\frac{1}{960}$&
 $\frac{121}{383040}$&
 -$\frac{71}{443520}$&
 $-\frac{19}{403200}$ \tabularnewline
\hline
\end{tabular}\end{center}

\caption{\label{cap:table}First 9 coefficients of the series
expansion (\ref{sol:powser}) calculated analytically. }
\end{table}

\begin{figure}
\centering
\includegraphics[width=0.75\columnwidth]{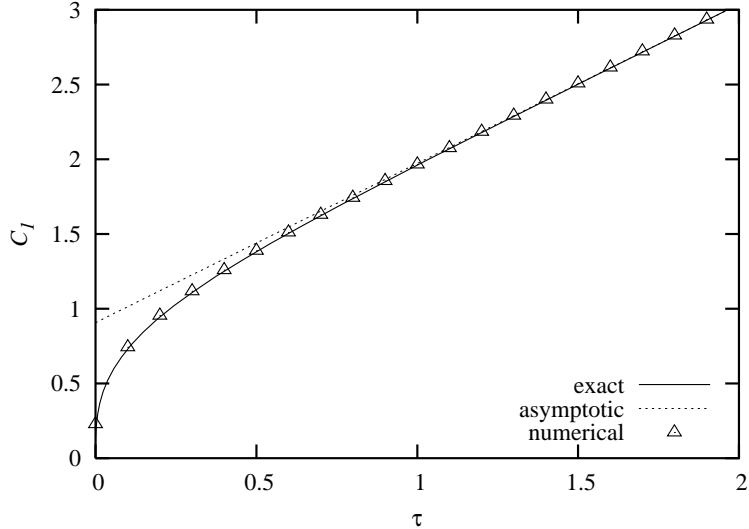}

\caption{
\label{cap:powser}
Solute distribution along the solidification front from the rim versus
$\tau=\ln(R/r)$ resulting from different approximations in comparison
to the numerical and exact solutions of Eq. (\ref{eq:C-tau}).}
\end{figure}

The Laplace transform yields also the asymptotic solution for
$\tau\gg1$ determined by the singularity of the image at $s=0$
where
\[
\frac{1}{s}F\left(\frac{s}{3},\frac{1}{3}\right)
\approx\frac{3}{\Gamma(1/3)}\frac{1}{s^{2}}\left(1+\frac{s}{3}
\left(\psi(1)-\psi\left(\frac{1}{3}\right)\right)\right)
\]
that straightforwardly leads to
\begin{equation}
C_{1}(r,0)\approx c_0
\left(\ln(R/r)+c_1\right),
\label{eq:C1r}
\end{equation}
where $c_{0}=\frac{3^{1/3}}{\Gamma\left(2/3\right)}\approx1.0651,$
$c_{1}=\frac{1}{3}\left(\psi\left(1\right)-\psi\left(\frac{1}{3}\right)\right)=
\ln\sqrt{3}+\frac{\pi}{6\sqrt{3}}\approx0.8516,$ and
$\psi(x)$ is the Psi (Digamma) function \cite{Abramowitz}.
This solution plotted versus $\tau=\ln(R/r)$ in Fig. \ref{cap:powser} is seen
to match both the numerical and the exact power series solution
(\ref{sol:powser}) surprisingly well already at $\tau>1$. The numerical
solution of Eq. (\ref{eq:C-tau}) is obtained by a Chebyshev collocation method
with an algebraic mapping to a semi-infinite domain for $z$ and a
Crank-Nicolson scheme for $\tau$ \cite{Canuto}.

\begin{figure}
\centering
\includegraphics[width=0.75\columnwidth]{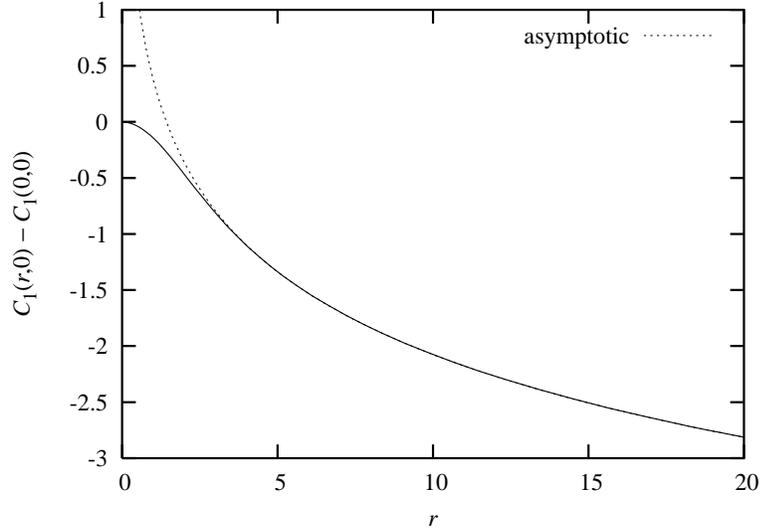}

\caption{\label{cap:cncola_r} Concentration perturbation relative
to its peak value (\ref{eq:C1-peak}) along the solidification
front in the vicinity of the symmetry axis together with the
corresponding asymptotic solution (\ref{eq:C1r}).}
\end{figure}

Note that the solution (\ref{eq:C1r}) describing the solute
concentration increasing along the solidification front as
$\sim\ln(R/r)$ is not applicable at the symmetry axis $r=0$ where
it becomes singular. This apparent singularity is due to the
neglected radial diffusion term in Eq. (\ref{eq:C-rbnd}) which,
obviously, becomes significant in the vicinity of the symmetry
axis at distances comparable to the characteristic thickness of
the solute boundary layer (\ref{eq:d0}) that corresponds to a
dimensionless radius of $r\sim 1.$ The radial diffusion becoming
effective at $r\lesssim1$ is expected to limit the concentration
peak at $\sim\ln(R).$ The asymptotic solution for the solute
boundary layer forming around the symmetry axis, which will be
published elsewhere because of its length and complexity, yields
for $R\gg1$ the peak value of the concentration perturbation at
the symmetry axis
\begin{equation}
C_{1}(0,0)\approx c_{0}(\ln(R)+c_{1})-c_{r},
\label{eq:C1-peak}
\end{equation}
where $c_{r}\approx0.3772.$ The concentration distribution along
the solidification front in the vicinity of the symmetry axis is
shown in Fig. \ref{cap:cncola_r}. As seen, the solution approaches
the finite value (\ref{eq:C1-peak}) at the symmetry axis while the
asymptotic solution (\ref{eq:C1r}) represents a good approximation
for $r\gtrsim 2.$ This solution is obtained numerically by a
Chebyshev collocation method \cite{Canuto} applied to Eq.
(\ref{eq:C-rbnd}) with the asymptotic boundary conditions
$\left.r\frac{\partial C_{1}}{\partial
r}\right|_{r\rightarrow\infty}= \left.z\frac{\partial
C_{1}}{\partial z}\right|_{z\rightarrow\infty}=-c_{0}$ supplied by
the outer asymptotic solution. This defines the solution in the
corner region at the symmetry axis up to an arbitrary constant
which is determined by matching with the outer analytic asymptotic
solution and yields the constant $c_r$ appearing in Eq.
(\ref{eq:C1-peak}). Note that in the described asymptotic
approximation the difference $C_1(r,0)-C_1(0,0)$ shown in Fig.
\ref{cap:cncola_r} is a function of $r$ only while the dependence
on $R$ is contained entirely in $C_1(0,0)$ defined by Eq.
(\ref{eq:C1-peak}).

\section{Summary and conclusions}

We have analysed the effect of a converging melt flow, which is
directed away from the solidification front, on the solute
distribution in several simple solidification models. First, it
was shown that the classical Burton-Prim-Slichter solution based
on the local boundary layer approach is not applicable for such
flows because of the divergence of the integral defining the
effective thickness of the solute boundary layer. Second, in order
to avoid this divergence we considered the model of a flat,
radially-unbounded layer of melt confined between two disks
representing melting and solidification fronts. This resulted in a
radially uniform solute distribution which, however, breaks down
as the velocity of the melt flow away from the solidification
front becomes comparable to the growth rate. This suggested that a
sufficiently strong radially converging melt flow is incompatible
with a radially uniform concentration distribution and,
consequently, radial solute segregation is unavoidable in such
flows. Thus, we next analysed the radial solute segregation caused
by a strong converging melt flow over a solidification front
modeled by a disk of finite radius $R_{0}$. We obtained an
analytic solution showing that the radial solute concentration at
the solidification front depends on the cylindrical radius $r$ as
$\sim\ln^{1/3}\left(R_{0}/r\right)$ and
$\sim\ln\left(R_{0}/r\right)$ close to the rim of the disk and at
large distances away from it, respectively. It is important to
note that these scalings do not imply any singularity at the axis
$r=0$. Instead, the concentration perturbation takes the value
(\ref{eq:C1-peak}) at the mid-point of the finite radius disk.

It has to be stressed that the radial segregation according to our
analysis is by a factor $\ln(R_{0}/d_{0})$ larger than that
suggested by a simple order-of-magnitude or dimensional analysis
(\textit{e. g.} Eq. (\ref{eq:C1def})). Thus, for converging flows
the concentration at the solidification front is determined not
only by the local velocity distribution but also by the ratio of
internal and external length scales which appear as a logarithmic
correction factor to the result of a corresponding scaling
analysis. The main conclusion is that flows converging along the
solidification front, conversely to the diverging ones, cause a
radial solute segregation with a logarithmic concentration peak at
the symmetry axis which might be an undesirable feature for
crystal growth applications.

\section{Acknowledgements}
Financial support from Deutsche Forschungsgemeinschaft in
framework of the Collaborative Research Centre SFB 609 and from
the European Commission under grant No. G1MA-CT-2002-04046 is
gratefully acknowledged.

\end{document}